\documentclass[aps,pre,twocolumn,showpacs,preprintnumbers,amsmath,amssymb]{revtex4-1}

\usepackage{dcolumn}
\usepackage{bm}

\usepackage[latin1]{inputenc}
\usepackage{amsfonts}
\usepackage{amsmath}
\usepackage{amssymb}
\usepackage{amsthm}
\usepackage{bbold}
\usepackage{color}
\usepackage[american]{babel}
\usepackage[T1]{fontenc}
\usepackage{longtable}
\usepackage[dvips]{graphicx}
\usepackage{xspace}
\usepackage{bbm}
\usepackage[all]{xy}

\newcolumntype{C}[1]{>{\centering\arraybackslash}p{#1}}

\begin{document}


\title{Nuclear Quantum Effects on the Vibrational Dynamics of Liquid Water}

\author{Deepak Ojha}
\author{Andr\'es Henao}
\affiliation{%
Dynamics of Condensed Matter and Center for Sustainable Systems Design, Department of Chemistry, University of Paderborn, Warburger Str. 100, D-33098 Paderborn, Germany
}
\author{Thomas D. K\"uhne}
\email{tdkuehne@mail.upb.de}
\affiliation{%
Dynamics of Condensed Matter and Center for Sustainable Systems Design, Department of Chemistry, University of Paderborn, Warburger Str. 100, D-33098 Paderborn, Germany
}
\affiliation{Paderborn Center for Parallel Computing and Institute for Lightweight Design, University of Paderborn, Warburger Str. 100, D-33098 Paderborn, Germany}

\date{\today}

\begin{abstract}
Based on quantum-mechanical path-integral molecular dynamics simulations the impact of nuclear quantum effects on the vibrational and hydrogen bond dynamics in liquid water is investigated. The instantaneous fluctuations in the frequencies of the O-H stretch modes are calculated  using the wavelet method of time series analysis, while the time scales of the vibrational spectral diffusion are determined from frequency-time correlation functions, joint probability distributions, as well as the slope of three-pulse photon echo. We find that the inclusion of nuclear quantum effects leads not only to a redshift of the vibrational frequency distribution by around 130~cm$^{-1}$, but also to an acceleration of the vibrational dynamics by as much as 30$\%$. In addition, quantum fluctuations also entail a significantly faster decay of correlation in the initial diffusive regime, which is agreement with recent vibrational echo experiments. 
\end{abstract}

\pacs{31.15.-p, 31.15.Ew, 71.15.-m, 71.15.Pd}
\keywords{}
\maketitle

\section{Introduction}

The significance of liquid water in several physical, chemical, biological, atmospheric and geophysical relevant processes has been extensively studied \cite{ball,ball2008, bb-book}. Many of its anomalous properties and rich phase diagram are believed to be a consequence of the 
temporally and spatially fluctuating hydrogen bond (HB) network, which itself is strongly correlated with the vibrational frequency of the O-H/O-D stretch modes present in liquid water \cite{ball, ball2008, bb-book, kauzmann}. Therefore, several non-linear spectroscopic experiments like three-pulse photon echo peak shift (3PEPS) and two-dimensional infrared spectroscopy (2D-IR) have been conducted to probe the rapid modulation of vibrational frequencies of the O-H/O-D modes \cite{zheng1, hamm, tokmak_1, tokmak_2, timmer, tokmak1, tokmak2, fayer1, chemrev}. Also, a large number of numerical simulations have analyzed the vibrational and HB dynamics using empirical and ab-initio interactions potentials \cite{skin2, skin3, hynes, hynes2, voth, saito, shin, ac, ac1, ac2, ac3, ohmine, tdk_water1, tdk_water3, tdk_water4, tdk_water5}. However, due to fact that water predominantly consists of light hydrogen atoms, the quantum-mechanical zero point energy (ZPE) and tunnelling effects have to be explicitly taken into account to yield the correct behavior of aqueous systems. The path-integral molecular dynamics (PIMD) method provides a particularly convenient way to efficiently incorporate nuclear quantum effects (NQE) within atomistic simulations \cite{hibbs, chandler, rahman, rossky, berne, haberson}.
Yet, although the importance of NQE in chemical processes involving hydrogen atoms, solvated protons, hydroxide ions and water itself have been duly acknowledged, the extent and magnitude of the impact on different static and dynamic properties is still ambiguous \cite{schmitt, haberson4, haberson5, lobaugh, tuckerman, spuraCC, ceriotti}. 

Recent experimental techniques such as deep inelastic neutron scattering (DINS) \cite{senesi1, senesi2, senesi3} allows to determine the proton and oxygen momentum distribution functions, as well as the quantum kinetic energies, which provide a direct estimation of NQE. Similarly, isotopic substitution effects on the structure, translational and orientational diffusion constants, dielectric relaxation times and viscosity have been looked at to estimate NQE in H$_{2}$O relative to D$_{2}$O \cite{lide, price, hardy, kauzmann}. For instance, it has been shown experimentally that the proton diffusion gets reduced by a factor of 1.5 upon deuteration \cite{roberts}. However, a recent computational study suggested the absence of NQE in the solvation of monovalent ions like Li$^{+}$ and F$^{-}$ \cite{wilkins}, whereas previous simulations have reported that the inclusion of NQE leads overall to a weakened HB network and consequently to a less structured liquid \cite{paesani1, rossky, tdk_water4, haberson3, paesani2, qRPC, laage}. 
In particular, earlier theoretical studies claimed that the inclusion of NQE leads to a faster diffusion of up to 50$\%$, while more recent PIMD simulations found that these effects are substantially smaller due to a sizable competition between inter- and intramolecular NQE \cite{paesani2, laage, tdk_water4, qRPC, haberson3}. 

However, the role of NQE on the vibrational dynamics of liquid water and especially the relation between the frequency-time correlation function (FTCF) and non-linear observables extracted from vibrational echo spectroscopy is as yet not well understood. Thus, in the present study, the instantaneous fluctuations of the ground state frequencies are computed using the wavelet method of time series analysis. Specifically, we determine the frequency fluctuation dynamics using the FTCF and the slope of three-pulse photon echo (S3PE) that is analogue to 3PEPS, which is calculated within the second order cumulant approximation. The computed spectral diffusion time scales are compared with a recent peak shift experimental study. Moreover, the HB lifetime and reorientation time scales, as well as the time-dependent frequency probability distribution, three-point frequency correlation and frequency structure correlation functions are computed. 

\section{Computational Details}

In the present work, we have performed PIMD simulations in the canonical NVT ensemble consisting of 216 water molecules using the flexible q-TIP4P/F force-field of Habershon and coworkers \cite{haberson3, tdk_water4, vrabec}.  However, at variance to modern many-body and machine-learning-based potentials \cite{xantheas, babin, medders, csanyi, behler}, the employed water model is neither polarizable, nor able to simulate chemical reactions that may take place in water \cite{chandler, hassanali1, hassanali2}. 
All of the simulations were conducted in a periodic cubic box of length $L=18.64$~\AA, which corresponds to the experimental density of liquid water at ambient conditions. 
Periodic boundary conditions were applied using the minimum image convention. Short-range interactions were truncated at 9~\AA \, and the Ewald summation technique was employed to calculate the long-range electrostatic interactions. The ring-polymer contraction scheme with a cutoff value of $\sigma=5$~\AA \, was used to reduce the electrostatic potential energy and force evaluations to a single Ewald sum, in order to significantly accelerate the calculations \cite{haberson3}. Specifically, $p=32$ ring-polymer beads were employed, while the computationally expensive part of the electrostatic interactions were contracted to just the centroid. Contrary to the original PIMD scheme, the partially adiabatic centroid MD technique permits to approximate dynamical quantities within the path-integral framework \cite{hone}. By modifying the elements of the Parrinello-Rahman mass matrix, the effective masses of the imaginary-time ring-polymer beads are chosen  so as to recover the correct dynamics of the centroids, while allowing for integration time-steps close to the ionic resonance limit. The temporal evolution of the ring-polymer was performed analytically in the normal mode representation by a multiple time-step algorithm using a discretized time-step of 0.5~fs for the intermolecular and 0.1~fs for the intramolecular interactions \cite{MTS}. Therewith, all systems were initially equilibrated for 20~ps followed by production runs of 100~ps each. For comparison, an additional simulation with classical nuclei with $p=1$ was performed. 



\section{Results and Discussion}

\subsection{Structure and Reorientational Dynamics}

\begin{figure}
\begin{center}
\includegraphics[width=0.475\textwidth]{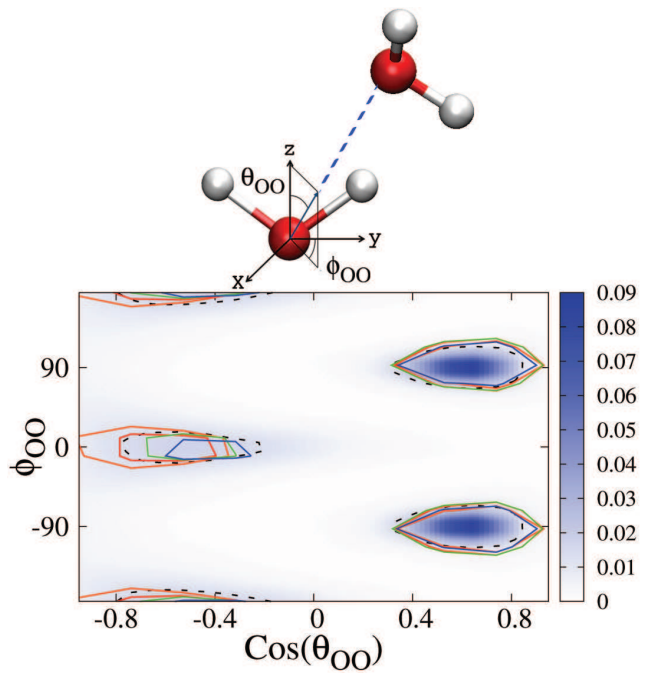}
\end{center}
\caption{\label{Fig1} Distribution function $g(\theta_{OO},\phi_{OO})$ describing the probability of finding a molecule in a certain region of space in spherical coordinates: dark blue color is associated with regions of high probability. The contour for $g(\theta_{OO},\phi_{OO})=0.01$ is shown for the first four neighbors (dashed line) together with that of the first (orange), second (red), third (green) and fourth neighbours (blue). The axis chosen to study the positions are exhibited at the top of the figure.
} \end{figure}

\begin{figure}
\begin{center}
\includegraphics[width=0.475\textwidth]{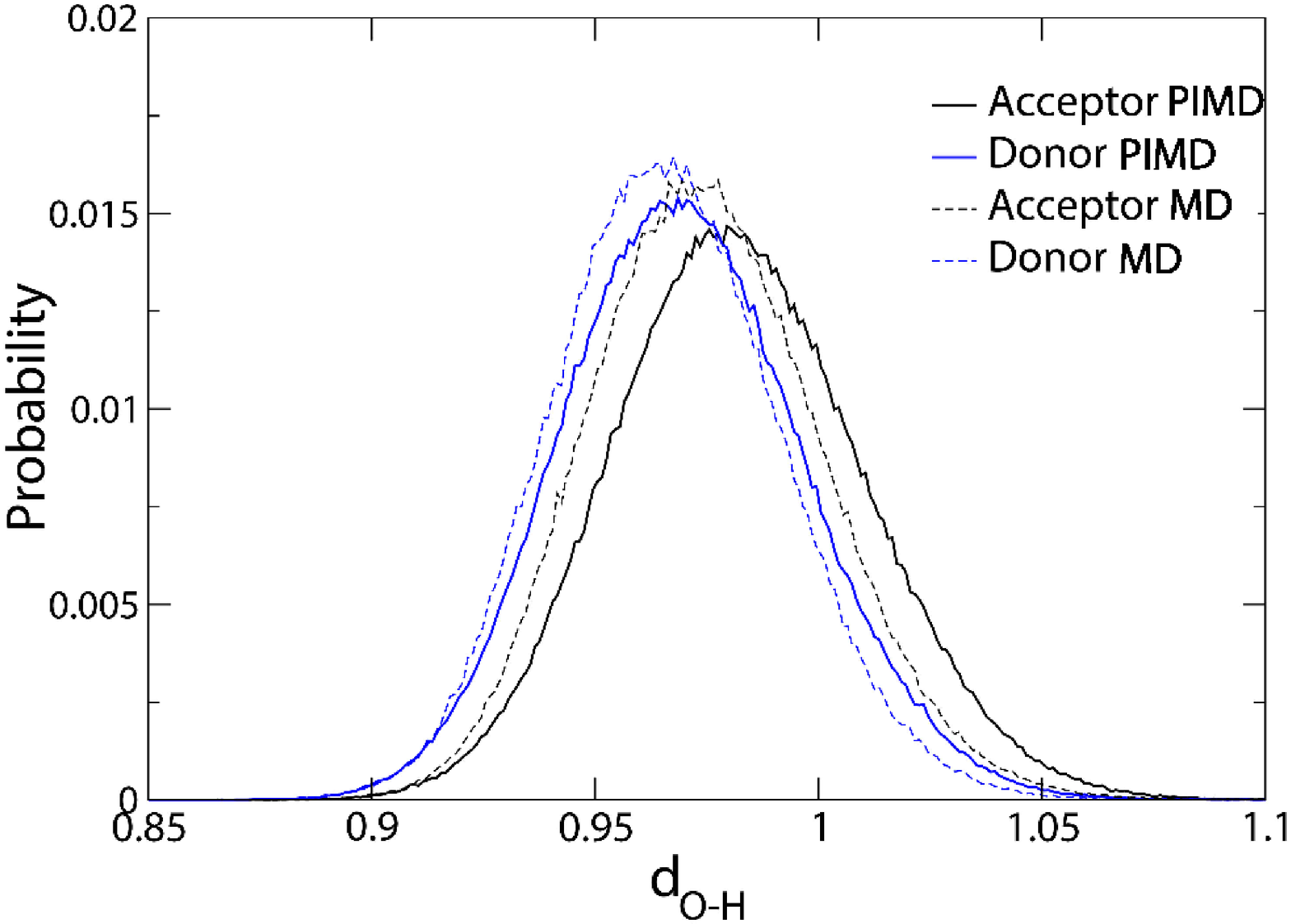}
\end{center}
\caption{\label{Fig2} Probability distribution functions for the intramolecular O-H bond length of the HB acceptor (black) and HB donor (blue) molecules. The continuous lines denote the results of the quantum-mechanical PIMD simulation, whereas the dashed lines corresponds to the classical MD simulation.
} \end{figure}

The local structure of liquid water is investigated using the angular distribution functions by associating an axis set to each molecule, where the Z axis is oriented along the dipole of each water molecule, whereas the X axis is set perpendicular to the H-O-H plane and the Y axis orthonormal to the X and Z directions  using the analysis program ANGULA \cite{Angula}, as illustrated in Fig.~\ref{Fig1}. The most predominant configurations are visualized in terms of the distribution function $g(\cos(\theta_\text{OO}),\phi_\text{OO})$ that describes the position of a neighboring oxygen atom with respect to the vector connecting both oxygen atoms, as shown in the inset of Fig.~\ref{Fig1}. The HB acceptor molecules are located in the northern ($\cos(\theta_\text{OO})>0$) and HB donors in the southern hemisphere ($\cos(\theta_\text{OO})<0$). 
The corresponding results, which are consistent with a tetrahedral structure of the first solvation shell, are presented in Fig.~\ref{Fig1}. 
Since the impact of NQE in this quantity is unnoticeable, the results of the classical MD simulation ($p=1$) are not shown explicitly. However, in Fig.~\ref{Fig2}, where the intramolecular O-H bond distance is shown, not only a clear distinction between the distances of HB donors and acceptors, but also a significant NQE can be detected. 
More precisely, the O-H bond length of the HB accepting molecules is throughout larger, which is a consequence of the previously recognized asymmetry between the HB donor and acceptor distances \cite{pardo, kuhne2013, tdk_water2}.  This asymmetry is consistent with the well-known fact that the distribution of electron acceptors around a water molecule is more disordered than that of the donors \cite{Agmon}. This phenomenon can be attributed to the existence of the so-called ``negativity track'' between the lone pairs of a water molecule, which facilitates the disordered motion of electron acceptors around the central donor \cite{Saykally, RustamJACS}
Furthermore, the inclusion of NQE entails a general increase in the average O-H bond length that again is consistent with previous results \cite{haberson3, tdk_water4}. 
\begin{figure}
\begin{center}
\includegraphics[width=0.475\textwidth]{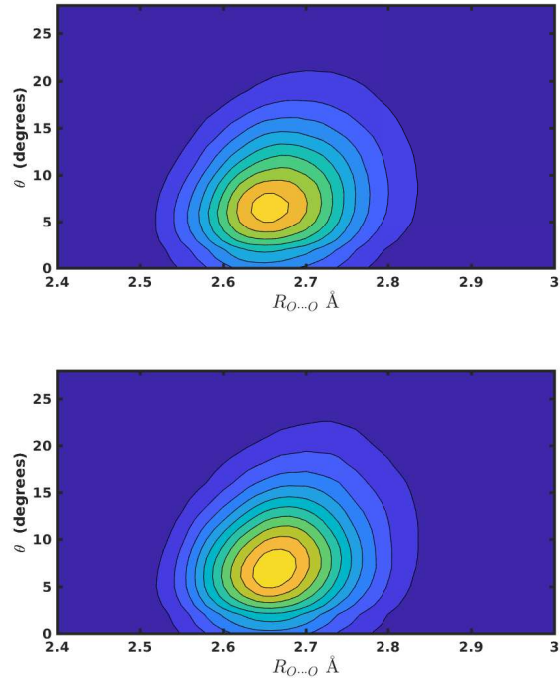}
\end{center}
\caption{\label{Fig3}
Intermolecular O$\cdots$O distance $R_{O \cdots O}$ as a function of the HB angle $\beta$ from classical MD (top panel) and PIMD (bottom panel) simulations. 
} \end{figure}
Looking at the HB angle  $\beta$ as a function of the intermolecular O$\cdots$O distance $R_{O \cdots O}$ (see Fig.~\ref{Fig3}), we find that the average distances are only slightly affected by NQE. 

For the purpose to study the reorientational dynamics of liquid water, we utilize the orientational time correlation function 
\begin{equation} \label{Eq1}
C_{2}^{\mu }(t) = \left \langle P_{2}\left [ \mu(0) \cdot \mu(t)  \right ] \right \rangle,
\end{equation}
where $\mu(t)$ is the orientation of the molecular dipole vector at time $t$ and $P_{2}$ the second-order Legendre polynomial.
\begin{figure}
\begin{center}
\includegraphics[width=0.475\textwidth]{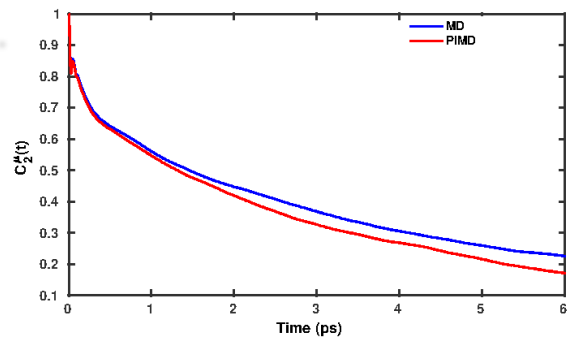}
\end{center}
\caption{\label{Fig4}
Orientational time correlation function of the molecular dipole vector of liquid water, as computed by MD and PIMD simulations.
} \end{figure}
The integrated time scale of the orientational time correlation function, which is shown in Fig.~\ref{Fig4}, gives the average time of the loss of correlation in the reorientational dynamics. For liquid water, the time scales of the orientational correlation function of the molecular dipole vectors is found to be 3.49~ps for the classical MD and 2.91~ps for the PIMD simulations, respectively. This is to say that the reorientational dynamics is accelerated due to NQE by as much as 17$\%$, which is in agreement with recent studies of others \cite{paesani2,laage}.

\subsection{Frequency Fluctuation Dynamics}

The instantaneous fluctuations of the ground state frequencies of the intramolecular O-H modes are computed by means of the wavelet method of time-series analysis, whose computational details are described elsewhere \cite{ac, ac2, wigg}. Here, we only briefly reiterate the underlying principles of the wavelet method that a time-dependent function can be expressed in terms of translations and dilations of a mother wavelet
\begin{equation} \label{Eq2}
\psi_{a,  b}(t)=a^{-\frac{1}{2}}\psi\Bigg(\frac{t-b}{a}\Bigg),
\end{equation}
where the coefficients of the wavelet expansion are given by the wavelet transform of $f(t)$, i.e. 
\begin{equation} \label{Eq3}
L_{\psi}f(a,  b)=a^{-\frac{1}{2}}\int_{-\infty}^{+\infty}f(t)\bar{\psi}
\Bigg(\frac{t-b}{a}\Bigg)dt.
\end{equation}
Herein, $a$ and $b$ are both real quantities with $a>0$. For the mother wavelet, the so-called Morlet-Grossman form is employed \cite{carmona}. Moreover, the inverse of the scale factor $a$ is proportional to the frequency over a time window around $b$. The time-dependent function $f(t)$ is constructed so as to be a complex function with its real and imaginary parts corresponding to the instantaneous bond length and momentum of a O-H mode projected along the O-H bond. This method is then applied to all the O-H modes present in a given system.

\begin{figure}
\begin{center}
\includegraphics[width=0.475\textwidth]{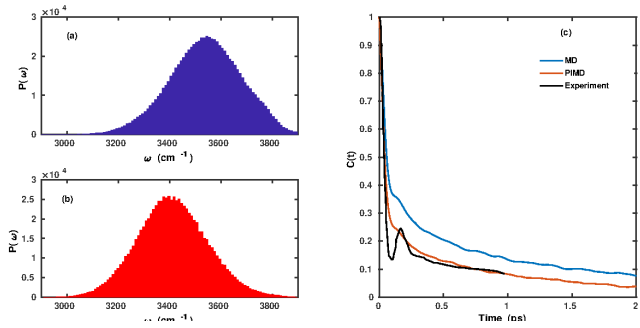}
\end{center}
\caption{\label{Fig5}
Vibrational frequency distributions of the O-H stretching modes of liquid water from (a) classical MD and (b) quantum-mechanical PIMD simulations. (c) Time-dependent decay of the FTCF as obtained by the MD and PIMD simulations, as well as experimental measurements of Ref.~\onlinecite{tokmak_3}. 
} \end{figure}
The distributions for the classical and quantum water systems are exhibited in Figs.~\ref{Fig5}(a) and \ref{Fig5}(b), respectively. 
The mean frequency of the O-H stretch oscillations as obtained by the classical MD and quantum-mechanical PIMD simulations are 3539~cm$^{-1}$ and 3412 cm$^{-1}$, respectively. Thus, the inclusion of NQE results in a redshift of 127~cm$^{-1}$. 
The FTCFs, which are shown in Fig.~\ref{Fig5}(c), captures the time scales on which the system loses its memory of the initial vibrational state and are defined as
\begin{equation} \label{EqFTCF}
C(t) = \left \langle \delta \omega(0) \cdot \delta \omega(t)  \right \rangle, 
\end{equation}
where $\delta \omega(t)$ is the instantaneous frequency fluctuation at time $t$. 
Moreover, the observables of multidimensional vibrational spectroscopies are also correlated with the system's FTCF. 
We see that the FTCF for both water systems exhibits a rather distinct decay pattern with a fast short-time decay in the first 100~fs, followed by a longer decay that extends up to a few picoseconds. For the purpose of extracting the time scales, the raw data of the MD and PIMD simulations are represented by a least-squares fit to the following bi-exponential function 
\begin{equation} \label{Eq4}
 \mathit{f(t)}= a_{0}exp\left(-\frac{t}{\tau _{0}}\right) + (1-a_{0})exp\left(-\frac{t}{\tau _{1}}\right), 
\end{equation}
where $\tau _{0}$ and $\tau _{1}$ are two characteristic time constants. The smaller of the two is generally attributed to the rearrangement of the HB network, while the larger is associated with the diffusive regime. 
The former is found to be 1.55~ps and 1.11~ps for the classical MD and PIMD simulations, respectively. Similarly, the dynamics in the diffusive regime is also accelerated due to NQE, which 
manifests itself in a large broadening of the time-dependent frequency probability distribution function of the PIMD simulation on rather short time scales within the sub-fs domain. While previous studies have shown that NQE can fasten the dynamics in liquid water by 15$\%$ \cite{laage, paesani2}, we find that the vibrational dynamics is accelerated by nearly 30$\%$. Moreover, we compare the computed FTCFs of the two simulations with the FTCF as obtained by mid-infrared spectroscopic measurements \cite{tokmak_3}. As is apparent from Fig.~\ref{Fig5}(c), the inclusion of NQE systematically improves the agreement with the experimental FTCF, both at short and long time scales.

\begin{figure}
\begin{center}
\includegraphics[width=0.475\textwidth]{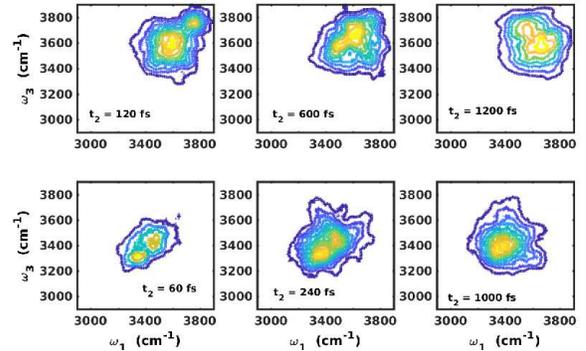}
\end{center}
\caption{\label{Fig6}
Joint probability distributions of finding the O-H stretching frequencies $\omega_{1}$ and $\omega_{3}$ of liquid water at a time delay of $t_{2}$ from MD (top) and PIMD (bottom) simulations.
} \end{figure}
The temporal evolution of the frequency fluctuations is analyzed by means of the joint probability distributions of finding the O-H stretching frequencies $\omega_1$ and $\omega_3$ separated by a time delay of $t_2$. It is shown in Fig.~\ref{Fig6} and enables to qualitatively determine the time scales of spectral diffusion. As a consequence, at short time delays, the distributions are localized around the diagonal, but gradually evolve into a completely delocalized spherical distribution within a time delay that corresponds to the time scales of the loss of frequency correlation. 
While this dephasing occurs for both water systems on qualitatively similar time scales, it is nevertheless happening faster and the distributions are generally slightly broader in the PIMD simulation. 

\begin{figure}
\begin{center}
\includegraphics[width=0.475\textwidth]{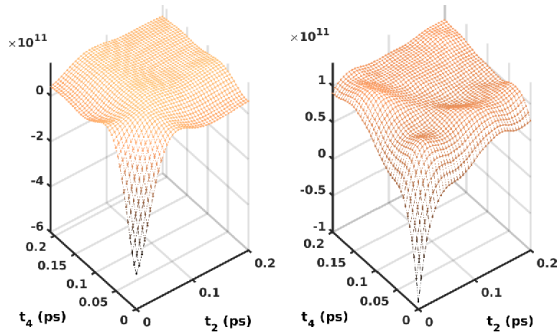}
\end{center}
\caption{\label{Fig7}
Time-dependent decay of the three-point correlation function as function of $t_{2}$ and $t_{4}$, as computed by MD (left panel) and PIMD (right panel) simulations.
} \end{figure}
Recent non-linear spectroscopic experiments such as 3D-IR spectroscopy have enabled the determination of three-point frequency correlation functions, which capture the differential dynamics of water molecules in different solvent environments \cite{hamm2}. 
In particular, it was demonstrated that these correlation functions obey initial oscillations with a change in sign, which can be attributed to the difference in the dynamics of the modes above and below the mean frequency of the absorption spectrum. Here, we have calculated the three-point frequency correlation functions, which are defined as
\begin{equation} \label{3PCF}
C(t_{2},t_{4}) = \left \langle \delta \omega(0) \cdot \delta \omega(t_{2}) \cdot \delta\omega(t_{2}+t_{4})  \right \rangle,
\end{equation}
based on our MD and PIMD simulations, as depicted in Fig.~\ref{Fig7}. We find that although initial short time scale oscillations can be observed for both cases, the oscillations are generally much more pronounced and extend over a longer duration of time if NQE are explicitly taken into account. This implies the existence of a differential dynamics of water molecules above and below of the mean frequency that is not short ranged.

\begin{figure}
\begin{center}
\includegraphics[width=0.475\textwidth]{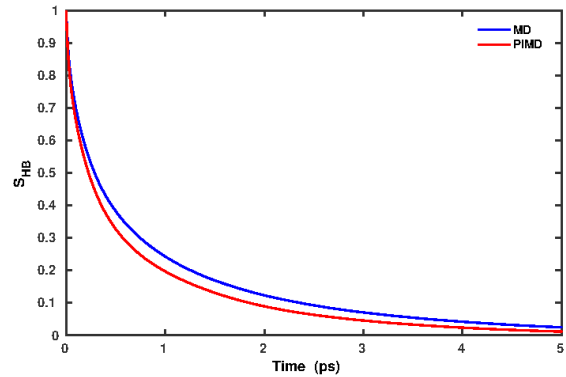}
\end{center}
\caption{\label{Fig8}
Time-dependent decay of the continuous HB correlation function of the molecular dipole vector of liquid water from classical and quantum-mechanical calculations. 
} \end{figure}
In order to elucidate the origin of the longer of the two time scales of the frequency correlation function, we have calculated the HB dynamics using a continuous HB correlation function approach \cite{dcr, luzar2, chandra1, chandra-pre}. Specifically, in Fig.~\ref{Fig8}, the HB time correlation function $S_{HB}(t)$, which denotes the probability that an initially H-bonded pair of water molecules remains continuously intact until time $t$, is presented for both systems. A HB between two water molecules is defined according to the criterium of Luzar and Chandler \cite{luzar1}. 
The HB lifetime $\tau_{HB}$ is set to the long-time decay constant $\tau_1$ as obtained by a least-squares fit to the bi-exponential function of Eq.~\ref{Eq4}. For the MD and PIMD simulations, $\tau_{HB}$ is 1.51~ps and 1.2~ps, respectively, which is in reasonable agreement with the slow decay component of the FTCF.
\begin{figure}
\begin{center}
\includegraphics[width=0.475\textwidth]{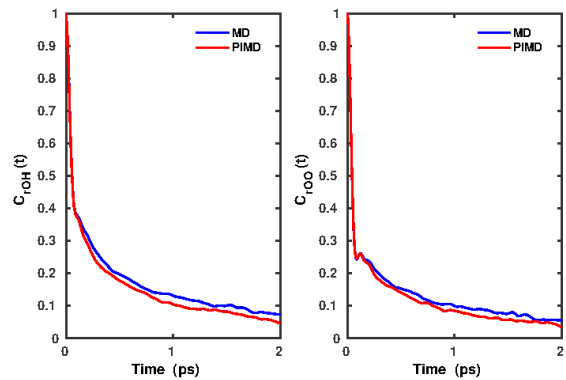}
\end{center}
\caption{\label{Fig9}
Time-dependent decay of the instantaneous fluctuations with respect to (a) $R_{O-H \cdots O}$ and (b) $R_{O \cdots O}$, as determined by MD and PIMD simulations.
} \end{figure}
Since the modulations in the vibrational frequencies of the O-H modes are influenced by its H-bonded partners, we calculate the time-dependent decay of fluctuations with respect to the $R_{H-O \cdots H}$ and $R_{O \cdots O}$ distances, as shown in Figs.~\ref{Fig9}(a) and \ref{Fig9}(b). The time scales of decay of the two correlation functions is again determined using a bi-exponential least-squares fit function  in analogy to Eq.~\ref{Eq4}. The fast decay component $\tau_0$ of the fit function extends up to a few hundred femtoseconds, whereas the slower decay component $\tau_1$ decays on a piccosecond time scale and is attributed to the loss of structural correlation. In the case of the classical MD simulation, the slower component of decay for $\left \langle \delta R_{H-O \cdots H}(0)\delta R_{H-O \cdots H}(t) \right \rangle$ is 1.38~ps and 1.36~ps for $\left \langle \delta R_{O \cdots O}(0)\delta R_{O\cdots O}(t) \right \rangle$. Adding NQE, the former reduces to 1.06~ps, while the latter amounts to 1.08~ps. A key feature in the decay of $\left \langle \delta R_{O \cdots O}(0)\delta R_{O \cdots O}(t) \right \rangle$ is a strong oscillatory behavior in the 100~fs regime, which has also been reported in previous experimental measurements \cite{tokmak_3}. In any case, this implies that the initial oscillatory pattern seen in the experimental FTCFs is predominantly due to the underdamped motion of intact H-bonded $O \cdots O$ pairs, whereas the long-term decay can be effectively described using either of the local structure parameters. 

\begin{figure}
\begin{center}
\includegraphics[width=0.475\textwidth]{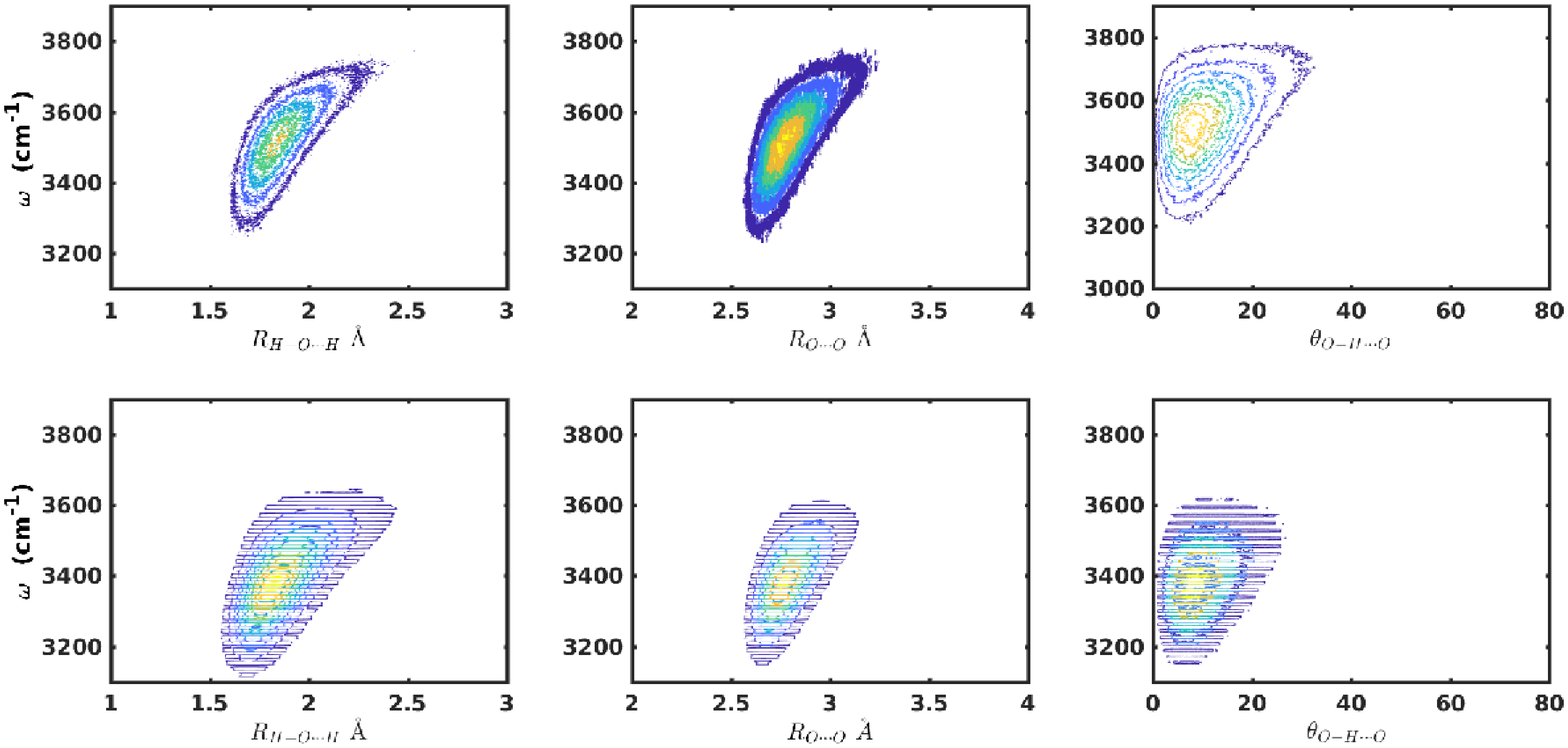}
\end{center}
\caption{\label{Fig10}
Frequency structure correlation between the instantaneous vibrational frequency of the O-H stretch mode as function of $R_{H-O \cdots H}$, $R_{O \cdots O}$ and the angle $\theta_{O-H \cdots O}$ from MD (top panel) and PIMD (bottom panel) simulations.
} \end{figure}
Along similar lines, the correlation between the vibrational frequencies and the spatial evolution of the local HB network can also be elucidated using the frequency structure correlation distributions. Specifically, in Fig.~\ref{Fig10}, the frequency structure correlation distributions of the instantaneous vibrational frequency as a function of $R_{O \cdots H}$, $R_{O \cdots O}$ and $\theta_{O-H \cdots O}$ are shown for the classical, as well as for the quantum water systems. We observe that the modulations in the vibrational frequencies the O-H modes are strongly influenced by its HB partners. Due to the fact that the inclusion of NQE results in a less structured liquid, it is not surprising that the results of the PIMD simulation possess a larger spreading of the vibrational frequencies as a function of $R_{O \cdots H}$ than the classical MD simulation. 


Non-linear spectroscopic techniques such as 3PEPS and 2D-IR have illustrated the ability to extract the frequency correlation loss. Within theoretical studies several analogues of the peak shift such as the S3PE have been proposed. It is obtained from the integrated echo intensity, which is given as
\begin{equation} \label{Eq5}
I(t_{1},t_{2}) =  \int_{0}^{\infty }dt_{3}\left |\sum_{i=1}^{3}R_{i}(t_{1},t_{2},t_{3})
\right |^2, 
\end{equation}
where the the response functions $R_{i}(t_{1},t_{2},t_{3})$ are calculated within the second order cumulant approximation \cite{mukamel-book,skinner2}. 
To extract the frequency correlation function from the integrated echo, we calculate the initial slope of the integrated three-pulse echo intensity, known as the short-time slope of the three-pulse echo or S3PE \cite{cho,skinner10,yang}, which can be mathematically expressed as
\begin{equation} \label{Eq6}
 S(t_{2})= \frac{\partial I(t_{1},t_{2}) }{\partial t_{1} }\mid _{t_{1}=0}.
\end{equation}
The normalized frequency-time correlation function is then given by 
\begin{equation} \label{Eq7}
 C(t_{2})= \frac{S(t_{2})}{S(0)}.
\end{equation}
\begin{figure}
\begin{center}
\includegraphics[width=0.475\textwidth]{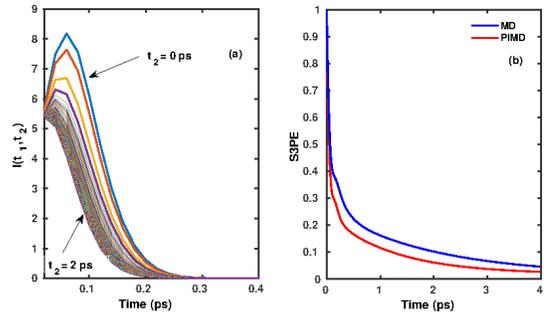}
\end{center}
\caption{\label{Fig12}
(a) Integrated echo intensity $I(t_{1},t_{2})$ as function of $t_{1}$, as obtained by PIMD. (b) Time-dependent decay of the S3PE from the classical and quantum-mechanical simulations.
} \end{figure}
The integrated echo intensity as a function of $t_{1}$ is given in Fig.~\ref{Fig12}(a), while the time-dependent decay of the S3PE as a function of $t$ is shown in Fig.~\ref{Fig12}(b). The time scales for the loss of correlation are computed using a least-squares fitting to a bi-expontential function as shown in Eq.~\ref{Eq1}. For the classical MD simulation, the slower of the two time constants is 1.87~ps, but reduces to 1.48~ps upon the inclusion of NQE. Even though the time scales are not in quantitative agreement with the afore reported time scales of the loss of memory on the initial vibrational state, the overall trend is essentially identical. Again, the inclusion of NQE leads to an acceleration of the vibrational dephasing rates by as much as 30$\%$. 

\section{Summary}

To summarize, we have in-depth studied the impact of NQE on the vibrational dynamics of liquid water using by means of PIMD simulations. We found that NQE generally reduces the time scales of frequency correlation by around 30$\%$. Moreover, apart from reducing the overall time scales of the loss of frequency correlation, these quantum fluctuations also entails a significantly faster decay of correlation in the initial diffusive regime. The results were found to be in reasonable agreement with previous vibrational echo experiments \cite{tokmak_3}. Furthermore, a redshift of the vibrational frequency distribution by approximately 130~cm$^{-1}$ is also observed. The trends seen in the decay of the vibrational correlation were reaffirmed by the calculation of the S3PE. While the computed time scales were not in quantitative agreement, they do predict an essentially identical trend of fastening the vibrational dynamics by as much as 30$\%$. 
The HB lifetime calculations also show that the average HB lifetime gets reduced by a similar factor by NQE. All of these results demonstrate that the inclusion of NQE are essential to understand the vibrational dynamics of liquid water in the initial diffusive, as well as long-tail decay regime. 

\begin{acknowledgments}
The authors would like to thank the Gauss Center for Supercomputing (GCS) for providing computing time through the John von Neumann Institute for Computing (NIC) on the GCS share of the supercomputer JUQUEEN at the J\"ulich Supercomputing Centre (JSC). This project has received funding from the European Research Council (ERC) under the European Union's Horizon 2020 research and innovation programme (grant agreement No 716142).
\end{acknowledgments}


\begin{thebibliography}{10}

\bibitem{ball}
P.~Ball,
\newblock {\em Life's Matrix: A Biography of Water}
\newblock (Univ of California Press 2001).

\bibitem{ball2008}
P. Ball,
\newblock {\em Chem. Rev.}  {\bf 108}, 74 (2008).



\bibitem{kauzmann}
D. S. Eisenberg and W. Kauzmann
\newblock {\em The Structure and Properties of Water}
\newblock  (Oxford,  1999).

\bibitem{bb-book} B. Bagchi,  Water in Biological and Chemical Processes: From Structure
and Dynamics to Function (Cambridge University Press 2013).


\bibitem{hamm}
J. Stenger, D. Madsen, P. Hamm, E. T. J. Nibbering and T. Elsaesser,
\newblock {\em  J. Phys. Chem. A} {\bf 106}, 2341 (2002).

\bibitem{zheng1}
T. Steinel, J. B. Asbury, J. Zheng and M.~D. Fayer,
\newblock {\em  J. Phys. Chem. A} {\bf 108}, 10957 (2004).

\bibitem{tokmak_1}
J.~J. Loparo, C.~J. Fecko, J.~D. Eaves, S.~T. Roberts and A.~Tokmakoff,
\newblock {\em Phys. Rev. B} {\bf 70}, 180201 (2004).

\bibitem{tokmak_2}
C. J. Fecko, J. J. Loparo, S. T. Roberts and A. Tokmakoff,
\newblock {\em  J. Chem. Phys.} {\bf 122}, 054506 (2005).

\bibitem{tokmak1}
J. J. Loparo, S. T. Roberts and A. Tokmakoff,
\newblock {\em  J. Chem. Phys.} {\bf 125}, 194521 (2006).

\bibitem{tokmak2}
J. J. Loparo, S. T. Roberts and A. Tokmakoff,
\newblock {\em J. Chem. Phys.} {\bf 125}, 194522 (2006).

\bibitem{timmer}
R.~L.~A. Timmer and H.~J. Bakker,
\newblock {\em J.  Phys. Chem. A} {\bf 113}, 6104 (2009).

\bibitem{fayer1}
M. D. Fayer,
\newblock {\em Acc.  Chem. Res.} {\bf 45}, 3 (2012).

\bibitem{chemrev}
F. Perakis, L. De Marco, A. Shalit, F. Tang, Z. R. Kann, T. D. K\"uhne, R. Torre, M. Bonn and Y. Nagata
\newblock {\em Chem. Rev.} {\bf 116}, 7590 (2016).

\bibitem{skin2}
C.~P. Lawrence and J.~L. Skinner,
\newblock {\em  J. Chem. Phys.} {\bf 118}, 264 (2003).

\bibitem{skin3}
Y.-S. Lin, B.~M. Auer and J.~L. Skinner,
\newblock {\em  J. Chem. Phys.} {\bf 131}, 144511 (2009).

\bibitem{ohmine}
M. Cho, G. R. Fleming, S. Saito, I. Ohimie and R.~M. Stratt,
\newblock {\em J. Chem. Phys.}  {\bf 100}, 6672 (1994).

\bibitem{hynes}
R. Rey, K.~B. Moller and J.~T. Hynes,
\newblock {\em  J.  Phys. Chem. A} {\bf 106}, 11993 (2002).

\bibitem{hynes2}
K. ~B. Moller, R. Rey and J.~T. Hynes,
\newblock {\em J.  Phys. Chem. A} {\bf108}, 1275 (2004).

\bibitem{ac}
B. ~S. Mallik, A.~Semparithi and A. Chandra,
\newblock {\em J.  Chem. Phys.} {\bf 129}, 19 (2008).

\bibitem{ac1}
 B. S. Mallik, A. Semparithi and  A. Chandra,
{\em J. Phys. Chem. A} {\bf 112}, 5104 (2008).

\bibitem{tdk_water1}
T. D. K\"uhne, M. Krack and M. Parrinello, 
\newblock {\em J. Chem. Theory Comput.} {\bf 5}, 235 (2009).

\bibitem{saito}
T. Yagasaki and S. Saito,
\newblock {\em Acc. Chem. Res.} {\bf 42}, 1250 (2009).

\bibitem{voth}
F. Paesani, S. ~S. Xantheas and G. ~A. Voth,
\newblock {\em J.  Phys. Chem. B}, {\bf 113}, 13118 (2009).

\bibitem{tdk_water3}
C. Zhang, R. Z. Khaliullin, D. Bovi, L. Guidoni and T. D. K{\"u}hne,
\newblock {\em J. Phys. Chem. Lett.} {\bf 4}, 3245 (2013).

\bibitem{shin}
S. Imoto, S. S Xantheas and S. Saito,
\newblock {\em J. Chem. Phys.}  {\bf  139}, 044503 (2013).

\bibitem{tdk_water4}
T. Spura, C. John, S. Habershon and T. D. K{\"u}hne,
\newblock {\em Mol. Phys.} {\bf 113}, 808 (2015).

\bibitem{ac2}
D. Ojha and A. Chandra,
\newblock {\em  J.  Phys. Chem. B} {\bf 119}, 11215 (2015).

\bibitem{tdk_water5}
C. Zhang, L. Guidoni and T. D. K{\"u}hne,
\newblock {\em J. Mol. Liq.} {\bf 205}, 42 (2015).

\bibitem{ac3}
D. Ojha and A. Chandra,
\newblock {\em  J.  Chem. Sci} {\bf 129}, 1069 (2017).

\bibitem{hibbs}
R. P. Feynman and A. R. Gibbs,
\newblock {\em Quantum Mechanics and Path Integrals}, {\em McGraw-Hill, New York} (1965).

\bibitem{chandler}
D. Chandler and P. G. Wolynes
\newblock {\em J.  Chem. Phys.} {\bf 74}, 4078 (1981).

\bibitem{rahman}
M. Parrinello and A. Rahman
\newblock {\em J. Chem. Phys} {\bf 80}, 860 (1984).

\bibitem{rossky}
R. A. Kuharski and P. J. Rossky,
\newblock {\em  J. Chem. Phys.} {\bf 82}, 5164 (1985).

\bibitem{berne}
A. Wallqvist and B. J. Berne,
\newblock {\em  Chem. Phys. Lett.} {\bf 117}, 214 (1985).


\bibitem{haberson}
S. Habershon, D. E. Manolopoulos, T.E. Markland and T. F. Miller,
\newblock {\em  Annu. Rev. Phys. Chem.} {\bf 64}, 387 (2013).



\bibitem{lobaugh}
J. Lobaugh and G. A. Voth,
\newblock {\em  J. Chem. Phys.} {\bf 104}, 2056 (1996).

\bibitem{schmitt}
U. W. Schmitt and G. A. Voth,
\newblock {\em  Chem. Phys. Lett.} {\bf 329}, 36 (2000).

\bibitem{tuckerman}
M. E. Tuckerman, D. Marx and M. Parrinello
\newblock {\em  Nature} {\bf 417}, 925 (2002).

\bibitem{haberson4}
 S. Habershon,
\newblock {\em  Phys. Chem. Chem. Phys.} {\bf 16}, 9154 (2014).

\bibitem{haberson5}
S. Habershon, G. S. Fanourgakis and D. E. Manolopoulos,
\newblock {\em  J. Chem. Phys.} {\bf 129}, 074501 (2008).

\bibitem{spuraCC}
T. Spura, H. Elgabarty and T. D. K{\"u}hne,
\newblock {\em Phys. Chem. Chem. Phys.} {\bf 17}, 14355 (2015).

\bibitem{ceriotti}
M. Ceriotti,  W. Fang, P. G. Kusalik, R. H. Mckenzie, A. Michaelides,
M. A. Morales and T. E. Markland,
\newblock {\em  Chem. Rev.} {\bf 116}, 7529 (2016).

\bibitem{senesi2}
 C. Andreani, G. Romanelli and R. Senesi,
\newblock {\em   Chem. Phys.} {\bf 427}, 106 (2013).

\bibitem{senesi1}
R. Senesi, G. Romanelli, M. A. Adams and C. Andreani,
\newblock {\em   Chem. Phys.} {\bf 427}, 111 (2013).

\bibitem{senesi3}
 G. Romanelli, M. Ceriotti, D.E. Manolopoulos, D. E. Pantelei,
R. Senesai and  C. Andreani,
\newblock {\em  J. Chem. Phys.} {\bf 139}, 074504 (2013).

\bibitem{lide}
D. R. Lide,
\newblock {\em CRC Handbook of Chemistry and Physics, 84th Edition}
\newblock  (Taylor \& Francis, 1999).

\bibitem{price}
W. S. Price, H. Ide, Y. Arata and  O Soderman,
\newblock {\em J. Phys. Chem. B} {\bf 104}, 5874 (2000).

\bibitem{hardy}
H. E. Hardy, A. Zygar, M. D. Zeidler, M. Holz and  F. D. Sacher,
\newblock {\em J. Chem. Phys.} {\bf 114}, 3174 (2001).

\bibitem{roberts}
N. K. Roberts and H. L. Northey,
\newblock {\em J. Chem. Soc.} {\bf 70}, 253 (1974).

\bibitem{wilkins}
D. M. Wilkins, D. E. Manolopoulos and  L. X. Dang
\newblock {\em J. Chem. Phys.} {\bf 142}, 064509 (2015).

\bibitem{paesani1}
F. Paesani, W. Zhang, D. A. Case, T. E. Cheetam and G. A. Voth,
\newblock {\em J. Chem. Phys.} {\bf 125}, 184507 (2006).

\bibitem{haberson3}
S. Habershon, T. E. Markland and D. E. Manolopoulos,
\newblock {\em  J. Chem. Phys.} {\bf 131}, 24501 (2009).

\bibitem{vrabec}
A. K\"oster, T. Spura, G. Rutkai, J. Kessler, H. Wiebeler, J. Vrabec and T. D. K\"uhne, 
\newblock {\em  J. Comp. Chem.} {\bf 37}, 1828 (2016).

\bibitem{xantheas}
G. S. Fanourgakis and S. S. Xantheas, 
\newblock {\em  J. Chem. Phys.} {\bf 128}, 074506 (2008).

\bibitem{babin}
V. Babin, G. R. Medders  and F. Paesani, 
\newblock {\em  J. Phys. Chem. Lett.} {\bf 3}, 3765 (2012).

\bibitem{medders}
G. R. Medders, V. Babin and F. Paesani, 
\newblock {\em  J. Chem. Theory Comput.} {\bf 9}, 1103 (2013).

\bibitem{csanyi}
A. P. Bartok, Michael J. Gillan, Frederick R. Manby and Gabor Csanyi, 
\newblock {\em  Phys. Rev. B} {\bf 88}, 054104 (2013).

\bibitem{behler}
T. Morawietz, A. Singraber, C. Dellago and J. Behler, 
\newblock {\em  Proc. Nat. Acad. Sci. USA} {\bf 113}, 8368 (2016).

\bibitem{chandler}
P. L. Geissler, C. Dellago, D. Chandler, J. Hutter and M. Parrinello, 
\newblock {\em  Science} {\bf 291}, 2121 (2001).

\bibitem{hassanali1}
A. Hassanali, M. K. Prakash, H. Eshet and M. Parrinello, 
\newblock {\em  Proc. Nat. Acad. Sci. USA} {\bf 108}, 20410 (2011).

\bibitem{hassanali2}
A. Hassanali, F. Giberti, J. Cuny, T. D. K\"uhne and M. Parrinello, 
\newblock {\em  Proc. Nat. Acad. Sci. USA} {\bf 110}, 13728 (2013).

\bibitem{paesani2}
F. Paesani, S. Yoo, H. J. Bakker and S. S. Xantheas,
\newblock {\em J. Phys. Chem. Lett.} {\bf 1}, 2316 (2010).

\bibitem{qRPC}
C. John, T. Spura, S. Habershon and T. D. K\"uhne, 
\newblock {\em J. Phys. Rev. E} {\bf 93}, 043305 (2016).

\bibitem{laage}
D. M. Wilkins, D. E. Manolopoulos, S. Pipolo, D. Laage and J. T. Hynes,
\newblock {\em J. Phys. Chem. Lett.} {\bf 8}, 2602 (2017).


\bibitem{hone}
T. D. Hone, P. J. Rossky and G. A. Voth,
\newblock {\em  J. Chem. Phys.} {\bf 124}, 154103 (2006).

\bibitem{MTS}
M. E. Tuckerman, B. J. Berne and G. J. Martyna,
\newblock {\em J. Chem. Phys.} {\bf 97}, 1990 (1992).

\bibitem{Angula}
\texttt{https://gcm.upc.edu/en/members/luis-carlos/angula/ANGULA}

\bibitem{pardo}
L. C. Pardo, A. Henao, S. Busch, E. Gu{\`a}rdia and  J. L. Tamarit,
\newblock {\em  Phys. Chem. Chem. Phys.} {\bf 16}, 24479 (2014).

\bibitem{kuhne2013}
T. D. K{\"u}hne and R. Z. Khaliullin,
\newblock {\em Nature Commun.} {\bf 4}, 1450 (2013).

\bibitem{tdk_water2}
R. Z. Khaliullin and T. D. K{\"u}hne,
\newblock {\em Phys. Chem. Chem. Phys.} {\bf 15}, 15746 (2013).

\bibitem{Agmon}
N. Agmon,
\newblock {\em Acc. Chem. Res.} {\bf 45}, 63 (2011).

\bibitem{Saykally}
F. N. Keutsch and R. J. Saykally,
\newblock {\em Proc. Nat. Acad. Sci. USA} {\bf 98}, 10533 (2001).

\bibitem{RustamJACS}
T. D. K\"uhne and R. Z. Khaliullin, 
\newblock {\em J. Am. Chem. Soc.} {\bf 136}, 3395 (2014).

\bibitem{wigg}
L. ~V. Vela-Arevalo and S. Wiggins,
\newblock {\em International Journal of Bifurcation and Chaos}
  {\bf11}, 1359  (2001).

\bibitem{carmona}
R. Carmona, W. Hwang and  B. Torresani, \textit{Practical Time-frequency
Analysis: Gabor and Wavelet Transforms with an Implementation in S},
Academic Press, 1998.

\bibitem{tokmak_3}
C.~J. Fecko, J.~D. Eaves, J.~J. Loparo, A.~Tokmakoff and P.~L. Geissler,
\newblock Ultrafast hydrogen-bond dynamics in the infrared spectroscopy of water.
\newblock {\em Science} {\bf 301}, 1698 (2003).

\bibitem{hamm2} S. Garrett and  P. Hamm,
{\em J. Chem. Phys.}  {\bf 128}, 104507 (2008).

\bibitem{dcr}
D.C. Rapaport,
\newblock {\em Mol. Phys.}  {\bf 50}, 1151, (1983).

\bibitem{luzar2}
A. Luzar, J. Chem. Phys. {\bf 113}, 10663 (2000).

\bibitem{chandra1} A. Chandra, Phys. Rev. Lett. {\bf 85}, 768 (2000)

\bibitem{chandra-pre} S. Chowdhuri and A. Chandra, Phys. Rev. E {\bf 66}, 041203 (2002).

\bibitem{luzar1}
A. Luzar and D. Chandler, Nature (London) {\bf 379}, 55 (1996).

\bibitem{mukamel-book}
S.~Mukamel,
\newblock {\em Principles Of Nonlinear Optical Spectroscopy}
\newblock (Oxford--New York, 1995).

\bibitem{skinner2}
J.~R. Schmidt, S.~A. Corcelli and J.~L. Skinner,
\newblock {\em  J.  Chem. Phys.}  {\bf 123}, 044513 (2005).


\bibitem{cho}
M. Cho, J.-Y. Yu; T. Joo, Y. Nagasawa, S. A.  Passino, G. R. Fleming,
\newblock {\em  J. Phys. Chem.} {\bf 100}, 11944 (1996).

\bibitem{skinner10}
K. F.  Everitt, E. Geva, J. L. Skinner,
\newblock  {\em  J. Chem. Phys.} {\bf 114}, 1326 (2001).

\bibitem{yang}
M.Yang,
 \newblock  {\em Chem. Phys. Lett.} {\bf 467}, 304 (2009).

\end{thebibliography}


\end{document}